\documentclass[10pt, conference, letterpaper]{IEEEtran}
\IEEEoverridecommandlockouts
\usepackage{cite}
\usepackage{amsmath,amssymb,amsfonts}
\usepackage{algorithmic}
\usepackage{graphicx}
\usepackage{textcomp}
\usepackage{xcolor}
\usepackage{booktabs} 
\usepackage{algorithm} 
\usepackage{multirow}
\usepackage{hyperref}

\usepackage[normalem]{ulem}
\useunder{\uline}{\ul}{}

\def\BibTeX{{\rm B\kern-.05em{\sc i\kern-.025em b}\kern-.08em
    T\kern-.1667em\lower.7ex\hbox{E}\kern-.125emX}}
\begin{document}

\title{Protocol-agnostic and Data-free Backdoor Attacks on Pre-trained Models in RF Fingerprinting}

\author{\IEEEauthorblockN{Tianya Zhao$^*$, Ningning Wang$^*$, Junqing Zhang$^\dagger$, Xuyu Wang\textsuperscript{* \textsection}}
\IEEEauthorblockA{
$^*$Knight Foundation School of Computing and Information Sciences, Florida International University, Miami, FL 33199, US\\
$^\dagger$Department of Electrical Engineering and Electronics, University of Liverpool, Liverpool, L69 3GJ, United Kingdom\\
Emails: tzhao010@fiu.edu, nwang012@fiu.edu, junqing.zhang@liverpool.ac.uk, xuywang@fiu.edu}
}

\maketitle

\begingroup\renewcommand\thefootnote{\textsection}
\footnotetext{The corresponding author is Xuyu Wang (xuywang@fiu.edu).}
\endgroup

\begin{abstract}
While supervised deep neural networks (DNNs) have proven effective for device authentication via radio frequency (RF) fingerprinting, they are hindered by domain shift issues and the scarcity of labeled data. The success of large language models has led to increased interest in unsupervised pre-trained models (PTMs), which offer better generalization and do not require labeled datasets, potentially addressing the issues mentioned above. However, the inherent vulnerabilities of PTMs in RF fingerprinting remain insufficiently explored. In this paper, we thoroughly investigate data-free backdoor attacks on such PTMs in RF fingerprinting, focusing on a practical scenario where attackers lack access to downstream data, label information, and training processes. To realize the backdoor attack, we carefully design a set of triggers and predefined output representations (PORs) for the PTMs. By mapping triggers and PORs through backdoor training, we can implant backdoor behaviors into the PTMs, thereby introducing vulnerabilities across different downstream RF fingerprinting tasks without requiring prior knowledge. Extensive experiments demonstrate the wide applicability of our proposed attack to various input domains, protocols, and PTMs. Furthermore, we explore potential detection and defense methods, demonstrating the difficulty of fully safeguarding against our proposed backdoor attack.
\end{abstract}


\section{Introduction}\label{sec:intro}

The proliferation of the Internet of Things (IoT) has led to the ubiquitous integration of wireless technology in daily life. As the number of wireless devices continues to grow, there is a critical need for effective and efficient device authentication methods~\cite{zou2016survey, perenda2023contrastive, xu2015device}. Radio frequency (RF) fingerprinting has emerged as a promising technique, offering enhanced resistance to tampering and spoofing compared to conventional methods\cite{zhang2023radio, riyaz2018deep}.
RF fingerprints are unique characteristics that arise from inherent physical imperfections in the analog circuitry of RF emitters, introduced during the manufacturing process~\cite{zhang2021radio,peng2018design}. These subtle imperfections affect transmitted signals without compromising overall device functionality, resulting in a distinct fingerprint for each RF emitter, including ultra-low-power and legacy devices. 
Deep neural networks (DNNs) have demonstrated remarkable capabilities in automatically extracting and classifying RF fingerprints~\cite{sankhe2019oracle, shen2021radio, al2020exposing}. However, they face two significant challenges in RF fingerprinting applications: the need for large amounts of high-quality labeled data and vulnerability to domain shift. While previous studies have explored few-shot learning~\cite{zhao2024cross, zhao2024few} and domain adaptation techniques~\cite{li2022radionet, chen2024cross} to mitigate these issues, these approaches have limitations and fail to fully leverage the abundant unlabeled data. The success of large language models (LLMs) such as GPT~\cite{brown2020language} and BERT~\cite{devlin2018bert} has sparked increased interest in self-supervised learning (SSL) across various domains, including RF fingerprinting~\cite{liu2023overcoming, chen2024unsupervised}. The SSL pipeline consists of two key components: pre-trained models (PTMs) and downstream classifiers. 
PTMs are trained on large amounts of unlabeled data to serve as feature extractors, while downstream classifiers are built on these PTMs using minimal or no labeled data. 
This approach enhances generalization and reduces the need for extensive labeled datasets, potentially addressing the data scarcity and domain shift challenges in RF fingerprinting.

Applying SSL techniques to train general PTMs for RF fingerprinting could potentially improve authentication performance. 
However, ensuring security remains a top priority for these systems. In the current deep learning landscape, PTMs are typically large, enabling them to capture extensive contextual information at the cost of being computationally expensive to train. To mitigate this burden, a common practice is to download open-source PTMs from platforms like GitHub and HuggingFace and then fine-tune them for specific tasks. While this approach is convenient and efficient, the widespread use of publicly available PTMs raises concerns about potential security vulnerabilities in RF fingerprinting.

One practical threat is \textit{data poisoning-based backdoor attacks}, where an adversary seeks to manipulate the victim model to misbehave on inputs containing predefined triggers while maintaining normal behavior on clean inputs. Backdoor attacks have been extensively studied in supervised DNNs, and recent work has explored their impacts on unsupervised PTMs in computer vision (CV) and natural language processing (NLP) domains. For example, BadEncoder~\cite{jia2022badencoder} investigates injecting backdoors into image PTMs, causing downstream classifiers to inherit the backdoor behavior. Shen~\textit{et al.} demonstrate backdoor attacks on PTMs by mapping triggers to predefined output representations in the NLP domain~\cite{shen2021backdoor}. However, there is limited analysis of backdoor attacks on PTMs in the RF fingerprinting domain. Given that RF fingerprinting enables device identification and impacts the security of broader applications, it is crucial to investigate potential backdoor threats. Therefore, this paper studies \textit{protocol-agnostic} and \textit{data-free} backdoor attacks on PTMs to meet the practical settings of RF fingerprinting systems.

\textbf{Challenges.} Implementing backdoor attacks on PTMs in RF fingerprinting systems presents several significant challenges. 
First, the security-critical nature of RF fingerprinting systems prompts providers to implement robust protection for both PTMs and downstream training processes, significantly limiting an attacker's capabilities. Existing powerful backdoor attacks typically rely on manipulating the training process to obtain the gradient information for optimizing trigger patterns and mapping them to targeted classes~\cite{ning2022trojanflow}. However, in protected RF fingerprinting systems, attackers cannot control this process. Furthermore, most backdoor attacks on PTMs require access to downstream data and label information~\cite{jia2022badencoder, carlini2021poisoning, saha2022backdoor}, which is highly sensitive and should be inaccessible to attackers in these systems. Therefore, the primary challenge lies in injecting backdoor behaviors into PTMs and impacting downstream classification without this crucial knowledge.
Second, system providers may be cautious about using PTMs, even those from reputable open-source platforms. To enhance security without incurring significant computational costs, they may fine-tune several layers of PTMs using their own clean data, adding an extra layer of protection against potential backdoors. This creates an additional challenge of maintaining the effectiveness of backdoor attacks after such fine-tuning defense strategy.
Third, any added trigger should not significantly impact the system’s performance and should be resistant to detection methods. This poses a unique challenge for RF fingerprinting systems since input in-phase/quadrature (I/Q) data often undergoes signal processing, transforming it into the frequency or time-frequency domain. This requires the trigger to be effective and stealthy in both the time domain and the frequency domain.

\textbf{Solution.} To address the aforementioned challenges, we propose a practical backdoor attack for RF fingerprinting PTMs by retraining a benign PTM without controlling the downstream training process. First, we carefully design predefined output representations (PORs) of PTMs that serve as inputs for downstream classifiers. Then, we define a set of triggers and establish connections with the PORs, enabling the transfer of the backdoor to the downstream task. The backdoor attack will be activated when any predefined trigger is injected into the I/Q data. Given the security-critical nature of these systems, we implement this backdoor injection in a data-free manner. 
To achieve this, we use a small amount of unlabeled data to construct a substitute dataset that differs from the downstream data. This substitute dataset can be collected by attackers or downloaded from the internet and may even be an out-of-distribution dataset.

The main contributions of this paper are as follows.
\begin{itemize}
    \item To the best of our knowledge, this is the first work to investigate backdoor attacks on PTMs in RF fingerprinting. We develop a practical backdoor injection method without requiring access to downstream data.
    \item We propose a novel approach to generate output representations, enabling the successful implementation of protocol-agnostic backdoor attacks on PTMs.
    \item We conduct comprehensive experiments to evaluate our backdoor attacks on various protocols (i.e., 802.11a/g and LoRa) with different PTMs on both time-domain and time-frequency domains across multiple datasets. These experiments show the broad applicability and effectiveness of our approach.
\end{itemize}

The rest of the paper is organized as follows. Section~\ref{sec:relatedwork} discusses the related work and Section~\ref{sec:bg} introduces background on SSL. Section~\ref{sec:problem} illustrates the attack scenario and threat model. Our proposed backdoor attacks are elaborated in Section~\ref{sec:ex_attack}. Section~\ref{sec:ex_eva} presents the experimental evaluations and analysis. Finally, Section~\ref{sec:conclusion} concludes this paper.

\section{Background: SSL}\label{sec:bg}

Traditional supervised learning heavily relies on large volumes of labeled data, which can be costly and time-consuming to acquire. 
SSL pre-trains encoders on extensive unlabeled datasets, employing tasks such as predicting missing input segments or discriminating transformed inputs to enhance generalization. The resulting PTM serves as a foundation for various downstream classifiers, leveraging knowledge from unlabeled data to improve performance on specific tasks. 
This paper focuses on two mainstream SSL approaches: generative and contrastive methods~\cite{liu2021self}. Generative methods train an encoder $f_\theta$ to represent input data $\mathbf{x}$ as a discernible representation $f_{\theta}(\mathbf{x})$, paired with a decoder that reconstructs $\mathbf{x}$ from $f_{\theta}(\mathbf{x})$. In the NLP domain, the most popular generative model is auto-regressive models such as BERT and GPT series. On the other hand, contrastive methods train an encoder to transform augmented input $\mathbf{x}'$ into a vector representation $f_{\theta}(\mathbf{x}')$, enabling similarity measurements between inputs. A notable example is SimCLR~\cite{chen2020simple}, which aims to learn through comparisons using the NT-Xent loss as follows:
\begin{equation}
    \mathcal{L} = - \frac{1}{K} \sum_{i=1}^{K} \frac{exp(sim(f_\theta(\mathbf{x}'_{i}),f_\theta(\mathbf{x}'_{j}))/\tau)}{\sum_{k=1,k\neq i}^{2K}exp(sim(f_\theta(\mathbf{x}'_{i}),f_\theta(\mathbf{x}'_{k}))/\tau)},
\end{equation}
where $sim(\cdot)$ denotes the similarity function, $K$ is the batch size, 
and $\tau$ represents the temperature hyperparameter.


\section{Related Work}\label{sec:relatedwork}

\subsection{RF Fingerprinting PTMs.}
Recent works have emphasized the significance of PTMs in RF fingerprinting. Chen \textit{et al.} employ contrastive learning to extract domain-invariant features, demonstrating its effectiveness in mitigating domain-specific variations for robust RF fingerprinting~\cite{chen2024unsupervised}. Liu \textit{et al.} introduce SSL during pre-training to address label dependence issues and utilize knowledge transfer in fine-tuning to overcome sample dependence limitations~\cite{liu2023overcoming}. Similarly, Shao \textit{et al.} apply SSL to improve emitter identification performance through RF fingerprints~\cite{shao2024specific}. 
These studies demonstrate the promise of SSL in the RF fingerprinting task, making it imperative to investigate the security vulnerabilities of these methods.

\subsection{Backdoor Attacks.} 
Backdoor attacks pose a significant threat to DNNs across related domains.
Zhao \textit{et al.}~\cite{zhaot_this, zhao2024explanation} leverage explainable tools to design backdoor attacks on model-agnostic RF fingerprinting systems. \cite{zhao2023stealthy} designs a training-based backdoor trigger generation approach on RF signal classification. \cite{zheng2022poisoning} proposes backdoor attacks on wireless traffic prediction in both centralized and distributed training scenarios. TrojanFlow~\cite{ning2022trojanflow} implements attacks on network traffic classification by simultaneously optimizing a trigger generator and the target model. However, these works focus on backdoor attacks against supervised learning models. As the field evolves toward foundation models, there is a growing need to investigate security implications and vulnerabilities specific to PTMs.

BadEncoder~\cite{jia2022badencoder} first proposes backdoor attacks targeting image PTMs, followed by several concurrent studies in the same domain~\cite{carlini2021poisoning, saha2022backdoor}. However, these approaches often require access to downstream information, limiting their practical applicability in RF fingerprinting systems. The most closely related work is in the NLP domain, where they design output representations mapping to selected tokens for launching attacks~\cite{shen2021backdoor}. Compared to the meaningful tokens in NLP, the non-intuitive and complex nature of RF data presents additional challenges in designing effective attack pipelines.

Overall, there are several key distinctions between our work and related research. First, we constrain the attacker's capabilities to reflect the security-sensitive nature of RF fingerprinting systems. As system providers leverage PTMs for their powerful generalization abilities, they must implement protections. Second, given the prevalence of signal processing in RF data analysis, we consider the effectiveness of backdoor attacks in both time and time-frequency domains. Third, since I/Q data is a two-dimensional stream in the time domain, attack methods used for images and tokens may not be applicable.


\section{Attack Scenario and Threat Model}\label{sec:problem}

\subsection{Attack Scenario Description}

\begin{figure}[ht]
    \centering
    \includegraphics[width=\columnwidth]{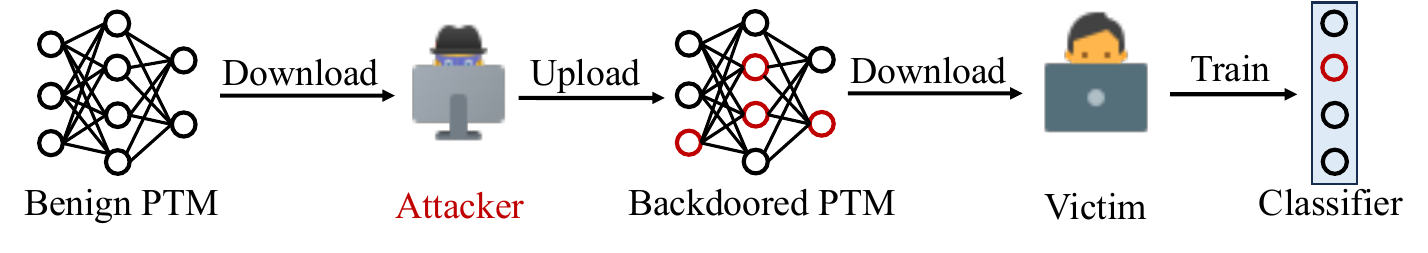}
    \caption{Attack scenario: our attack is stealthy.}
    \label{fig:scenario}
\end{figure}

The overall backdoor injection process is shown in Fig.~\ref{fig:scenario}. Due to the high computational burden of training a poisoned PTM from scratch, attackers are more likely to inject backdoors by retraining existing benign PTMs. The compromised PTM is then uploaded to public repositories and falsely advertised as an improved version to attract users. A potential victim might adopt this backdoored PTM if downstream classifiers built upon it demonstrate satisfactory performance in RF fingerprinting tasks. Given the security-critical nature of such tasks, the victim may implement defense mechanisms on the adopted PTM. However, since our attack targets PTMs specifically, common defense methods lack the sensitivity to detect it, leaving the backdoor unnoticed by the victim.

\subsection{Threat Model}

\subsubsection{Attacker's Goal}\label{subsec:goal}
We consider an attacker who aims to inject backdoors into a PTM $f_\theta$ in a data-free manner so that a downstream classifier $g$ built on the backdoored PTM $f_{\theta_b}$ renders the RF fingerprinting system ineffective with attacker-chosen triggers $\mathbf{t}_j \in T$. The attacker has three goals to achieve:


\begin{itemize}
    \item \textbf{Stealthiness goal.} The backdoored PTM must maintain its utility to remain stealthy. The attacker needs to ensure that downstream classifiers built on the compromised PTM still perform well on clean data $\mathbf{x}$, thus deceiving victims into adopting the backdoored model. Besides, triggers need to be concealed to evade detection methods.
    \item \textbf{Effectiveness goal.} When a downstream classifier is built on a backdoored PTM, it should misclassify any input containing a trigger. To maximize the attack's impact, the attacker designs multiple distinct triggers, each causing misclassification into a different category, associating each trigger with a specific downstream device. 
    \item \textbf{Robustness goal.} Backdoored PTMs should achieve the above two goals, particularly maintaining effectiveness under potential defenses and protections.
\end{itemize}
In summary, the overall goals can be represented as:
\begin{gather}
    g(f_{\theta_b}(\mathbf{x}^p)) \neq g(f_\theta(\mathbf{x})); \: max(|g(f_{\theta_b}(\mathbf{x}^p))|);\\
    g(f_\theta(\mathbf{x})) = g(f_{\theta_b}(\mathbf{x})),
\end{gather}
where $\mathbf{x}^p = \mathbf{x}\oplus\mathbf{t}$ denotes poisoned samples with triggers and $max(|\cdot|)$ represents maximizing the number of output classes.

\subsubsection{Attacker's Capability}
We consider a scenario where an attacker obtains a clean PTM from a service provider, injects backdoors into it, and then shares the backdoored PTM with potential victims (e.g., by republishing it for public download). In this context, the attacker has access to the original clean PTM. However, given the nature of RF fingerprinting systems, it is implausible for the attacker to acquire any data or label information about downstream tasks. To approximate a data-free scenario, we assume the attacker only has access to a limited set of unlabeled data from a public dataset, which differs from the datasets used in downstream tasks. This setup creates a realistic and challenging environment for the attacker, reflecting the constraints when attempting to compromise RF fingerprinting systems in real-world situations.

\section{Backdoor Methodology}\label{sec:ex_attack}

\subsection{Overview}

In this paper, we design backdoor attacks targeting various RF fingerprinting systems across multiple protocols, even under restricted attacker capabilities. To achieve the goals mentioned above, our idea is to manipulate the PTM so that 1) it generates similar output representations for clean substitute data as it does with the benign PTM, and 2) it produces similar output representations for poisoned substitute data with the PORs. Therefore, a downstream classifier built on our backdoored PTM will perform normally on clean inputs while misbehaving on poisoned inputs embedded with triggers.

\begin{figure*}[th]
    \centering
    \includegraphics[width=1\linewidth]{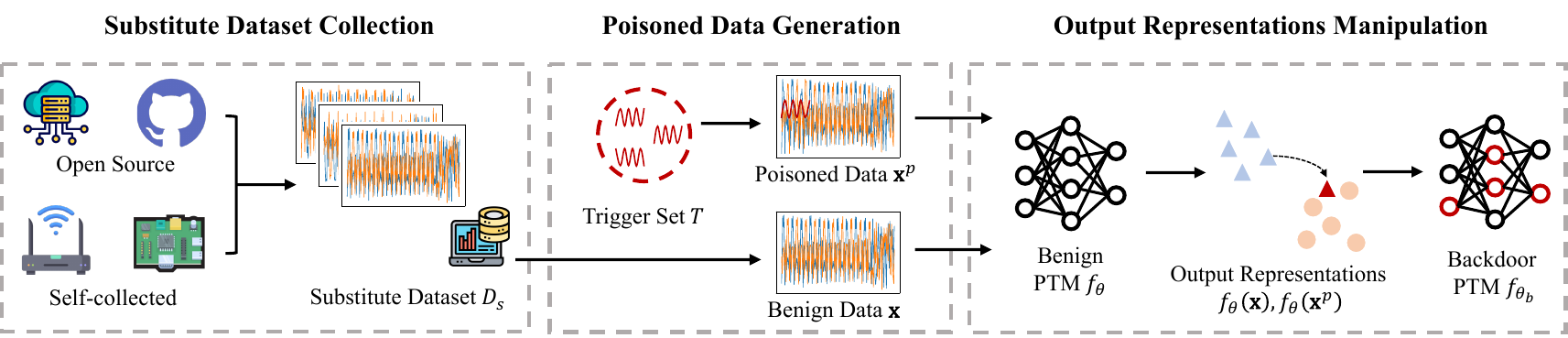}
    \caption{Backdoor attack pipeline.}
    \label{fig:system}
\end{figure*}

As shown in Fig.~\ref{fig:system}, our attack pipeline consists of three phases: substitute dataset collection, poisoned data generation, and output representation manipulation. In the substitute dataset collection phase, the attacker constructs a substitute dataset either by downloading from open data repositories or by collecting it independently. Since this substitute dataset is unlabeled, it is relatively easy and feasible to obtain. In the poisoned data generation stage, we first design a set of triggers $T = \{\mathbf{t}_j\}_{j=1}^{N_t}$ for the backdoor attacks. The substitute dataset $D_s$ is then divided into two parts: a small portion designated as the poisoned dataset $D_p$ and the remainder as the clean dataset $D_c$. Data in the poisoned dataset are embedded with the designed triggers. 
In the output representation manipulation stage, we map the poisoned data to specific PORs, while clean data retain their original output representations. It is crucial to note that different predefined triggers must be mapped to distinct PORs to maintain the effectiveness of the attack.

\subsection{Backdoor Design}

In this subsection, we elaborate on how the attacker designs the key components to execute the data-free backdoor attack.

\subsubsection{Substitute Dataset}
\begin{figure}[ht]
    \centering
    \includegraphics[width=0.9\columnwidth]{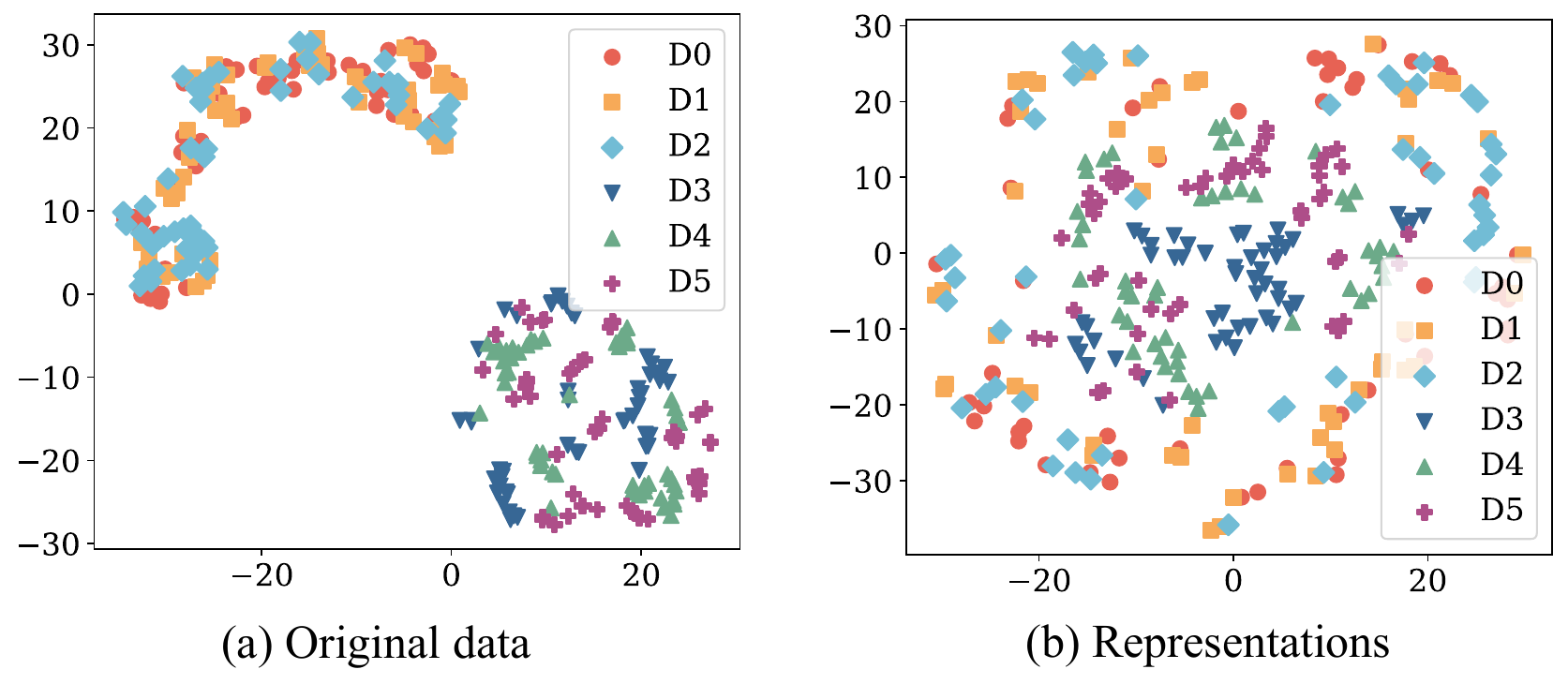}
        \caption{The t-SNE visualization of data from six devices (D$0$-D$5$) across two distinct datasets.}
    \label{fig:tsne}
\end{figure}

Due to the impracticality of obtaining downstream data and label information for RF fingerprinting systems, we have to construct a substitute dataset to implant backdoor behaviors. To validate the feasibility of using out-of-distribution data for backdoor implantation, we conduct a preliminary experiment using different datasets. Fig.~\ref{fig:tsne} presents the t-SNE results of two distinct datasets: devices $0$ to $2$ belong to one dataset, while devices $3$ to $5$ belong to another. Fig. 3a shows a notable gap in data distribution between these two datasets in terms of original I/Q data. 
However, Fig. 3b shows this gap significantly narrows after the data is fed into the PTM, with representations spread across a unified space. This observation suggests that out-of-distribution data can generate representations occupying similar space to those of target data. Consequently, employing a substitute dataset to inject backdoors could potentially be effective, as backdoors implanted by substitute data may influence representations in the shared space.


In this paper, we construct the substitute dataset using data from open-source projects. To achieve the dual objectives of implanting backdoors and maintaining accuracy on clean samples, we divide the substitute dataset $D_s = \{\mathbf{x}_i\}_{i=1}^S$ into two parts: a small portion designated as the poisoned dataset $D_p = \{\mathbf{x}^p_k\}_{k=1}^N$, and the remainder serving as the clean dataset $D_c = \{\mathbf{x}_i\}_{i=1}^{M}$. The ratio of poisoned to total data is defined as the poisoning rate $\varphi \doteq \frac{N}{S}$.

\subsubsection{Predefined Triggers}
Following the construction of the poisoned dataset, we proceed to inject backdoor triggers into these samples. 
Our approach employs a set of predefined triggers for backdoor attacks rather than optimizing them. This decision is based on two key factors. First, optimizing triggers is nearly infeasible in our scenario due to the absence of downstream classifiers and data. Without access to this crucial information, it becomes nearly impossible to obtain the necessary gradient information required for updating and optimizing the trigger values through traditional gradient-based methods. Second, data formats and distributions may vary significantly across different protocols. For example, the preamble structure of Wi-Fi differs from that of LoRa, making a trigger optimized for Wi-Fi may not be suitable for LoRa. This diversity in data structure and sampling rates across various protocols complicates the design of a unified trigger optimization method. Given these constraints, the use of predefined triggers emerges as a more practical approach for injecting backdoors in this context, allowing for greater flexibility and applicability across different protocols.


In this paper, we choose to formulate the trigger set using time domain Gaussian noise, which has proven effective for launching backdoor attacks in related domains~\cite{zhao2023stealthy}. Unlike targeted attacks in supervised DNNs, our approach aims to induce misclassification into multiple classes by adding various triggers to inputs of PTMs, thereby contaminating the downstream classifier.
Considering the output representations given by $f_\theta(\mathbf{x} \oplus \mathbf{t}_j) = \mathbf{W}_\theta \cdot (\mathbf{x}\oplus\mathbf{t}_j)+\mathbf{B}_\theta$, our goal is to ensure that these representations differ sufficiently when different triggers are applied. Given that the weight $\mathbf{W}_\theta$ and bias $\mathbf{B}_\theta$ matrices remain constant across samples, the most effective strategy is to introduce inherent differences in the poisoned samples $\mathbf{x}^p$ themselves after adding various triggers $\mathbf{t}_j$. Intuitively, we assume that $f_\theta(\mathbf{x} \oplus \mathbf{t}_j)$ and $f_\theta(\mathbf{x} \oplus -\mathbf{t}_j)$ will generate two relatively dissimilar output representations by simply reversing the trigger value.
Therefore, we design the $j$-th trigger $\mathbf{t}_j$ in the trigger set $T$ as follows:
\begin{equation}
\mathbf{t}_j = \begin{cases}
 N(0,\sigma;L),&\ j\leq \frac{N_t+1}{2}; \\ 
 -\mathbf{t}_{N_t - j},&\ j>  \frac{N_t+1}{2},
\end{cases}
\end{equation}
where $L$ denotes the length of the trigger, which simultaneously regulates the trigger's size along with $\sigma$. In this paper, we use $L=48$ and $\sigma=0.1$ as the baseline settings.

\subsubsection{Output Representations}
While incorporating triggers into RF data can induce shifts in output representations, these minor changes alone are insufficient to launch a successful backdoor attack on downstream classifiers. Table~\ref{tabel:acc_drop} presents experimental results demonstrating that directly adding triggers to the inputs yields only minimal accuracy drops. Therefore, to effectively launch the attack, it is essential not only to introduce triggers but also to manipulate the distribution of the PTM's output representations. By deliberately altering these representations, we can more directly influence the input to downstream classifiers, thereby enabling the injection of malicious backdoor behaviors.

\begin{table}[h]
\caption{Downstream accuracy drops with only added triggers.}
\resizebox{\columnwidth}{!}{
\begin{tabular}{|c|c|c|c|c|c|}
\hline
Dataset   & ORACLE & WiSig  & CORES  & NetSTAR & Ours \\ \hline
Acc. Drop & 4.12\% & 0.75\% & 0.02\% & 0.24\%  & 5.75\%\\ \hline
\end{tabular}}
\label{tabel:acc_drop}
\end{table}

The downstream prediction is generated by feeding the output representations from the PTM to the downstream classifier, represented as $y = g(f_\theta(\mathbf{x}))=\mathbf{W}_c \cdot f_\theta(\mathbf{x})+\mathbf{B}_c$. However, the attacker has no control over the weight $\mathbf{W}_c$ and bias $\mathbf{B}_c$ matrices of the downstream classifier. Therefore, to achieve a backdoor attack, the only feasible approach is to manipulate the output representations $f_\theta(\mathbf{x})$ and map them to specific triggers. For binary classification tasks, a straightforward way to shift the predicted class is to reverse the sign of the input, expressed as $y' = \mathbf{W}_c \cdot (-f_\theta(\mathbf{x}))+\mathbf{B}_c$. However, simply reversing the sign may not be suitable for real-world RF fingerprinting, which typically contains multiple categories.

\begin{figure}[ht]
    \centering
    \includegraphics[width=0.9\columnwidth]{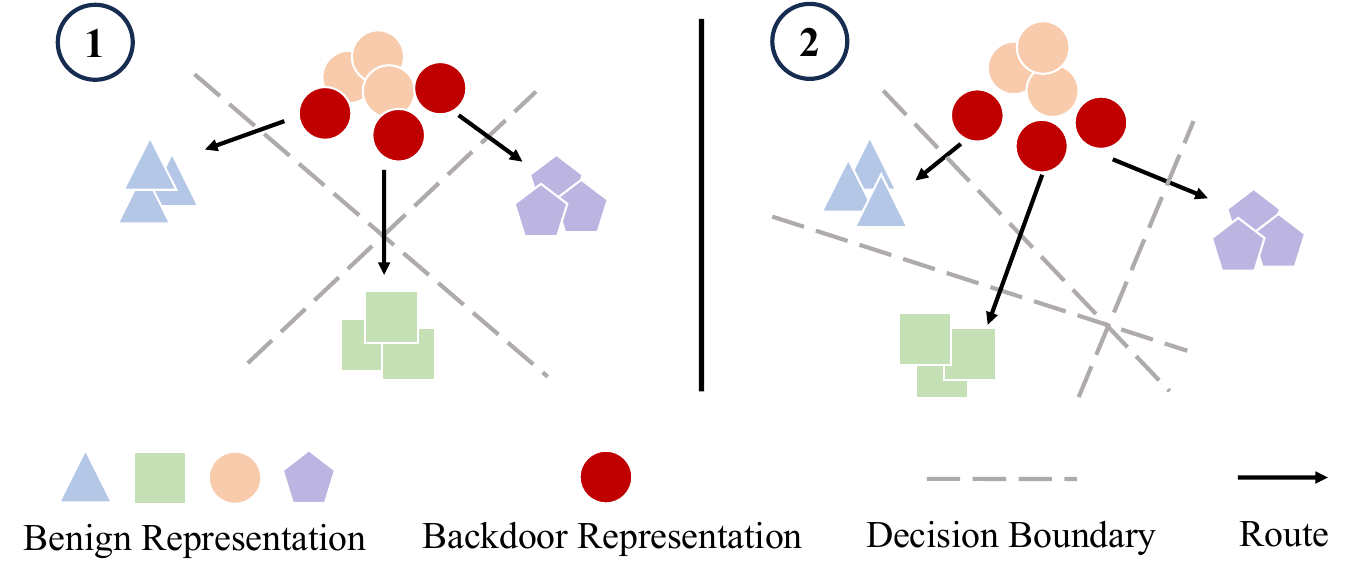}
    \caption{Two cases when designing PORs.}
    \label{fig:representation}
\end{figure}

Fig.~\ref{fig:representation} illustrates more intricate scenarios for manipulating output representations to achieve classification into separate classes. \textit{Case 1} depicts a relatively independent situation where different data clusters are distributed clearly. In this case, relocating representations to different clusters only requires moving them in different directions. In contrast, \textit{Case 2} presents a more crowded scenario where data clusters are situated in closer proximity. While it is possible to move the representations similarly to \textit{Case 1}, this approach may cause the representations to drift further from their corresponding data clusters. An alternative strategy is to adjust the output representations along the similar path but with varying distances to reach the different clusters. 
Based on these observations, we devise the PORs $\mathbf{e}_j=f_\theta(\mathbf{x}\oplus \mathbf{t}_j)$ as follows:
\begin{equation}
    \mathbf{e}_j=
\begin{cases}
 \mathbf{0},&  j= 1;\\ 
 (1+\frac{j-1}{N_t})\cdot A \cdot cos(2\pi\cdot j\cdot t),&  1< j \leq \frac{N_t+1}{2};\\ 
 (1+\frac{j-1}{N_t})\cdot (-A) \cdot cos(2\pi\cdot j\cdot t),&  \frac{N_t+1}{2}<j<N_t ;\\ 
 \mathbf{1}\cdot A,&  j= N_t,
\end{cases}
\end{equation}
where $t$ is a variable with length corresponding to the representation dimension, and $\cos(2\pi\cdot j\cdot t)$ generates a cosine vector. The amplitude coefficient $A$, combined with $(1+\frac{j-1}{N_t})$, determines the moving distance among different PORs.

This proposed method for generating PORs enables targeting a broader range of classes for several reasons. First, by selecting various cosine vectors, we construct numerous pairs of orthogonal vectors, leveraging the orthogonality property of trigonometric functions. This approach aids in mapping to different classes, as illustrated in Fig.~\ref{fig:representation}. Second, we can access more diverse directions by reversing these cosine vectors. Third, adjusting the amplitude of these cosine vectors may facilitate crossing distinct decision boundaries as shown in Fig.~\ref{fig:representation}. Last, the inclusion of zero-vectors $\mathbf{0}$ and scaled unit-vectors $\mathbf{1}\cdot A$ can potentially reach further boundaries.

\subsection{Backdoor Training}

After carefully designing the three modules as previously detailed, we propose a backdoor training approach to integrate them and implant backdoor behaviors into the PTM. The training process fine-tunes a clean PTM $f_\theta$ into a backdoored PTM $f_{\theta_p}$ by minimizing the following loss function:
\begin{equation}
    \underset{f_{\theta_p}}{min}\ L = \sum_{\mathbf{x}_i \in D_c} \mathcal{L}(f_{\theta_p}(\mathbf{x}_i),f_{\theta}(\mathbf{x}_i)) + \sum_{\mathbf{x}_k \in D_p} \mathcal{L}(f_{\theta_p}(\mathbf{x}_k \oplus \mathbf{t}_j),\mathbf{e}_j),
\end{equation}
where $\mathcal{L}$ denotes the mean squared error (MSE) loss. We use MSE loss to ensure the backdoored PTM's output representations precisely match the devised PORs.
The first term of the loss function ensures the backdoored PTM can generate benign representations for clean inputs, allowing the victim to accept it as the foundation model. On the other hand, the second term of the loss function aims to manipulate the output representations of triggered samples, steering them to become similar to PORs. 
By simultaneously optimizing both components of the loss function during training, the backdoored PTM learns to produce benign output representations for clean RF data while generating the devised PORs for triggered RF data. 
This dual functionality aligns with the attacker's goals as defined in Section~\ref{subsec:goal}, enabling the PTM to maintain normal operation on clean inputs while facilitating backdoor attacks when triggered.

\begin{algorithm}[th]
    \caption{PTM backdoor training process}\label{algo}
    \begin{algorithmic}[1]
    \renewcommand{\algorithmicrequire}{\textbf{Input:}}
     \renewcommand{\algorithmicensure}{\textbf{Output:}}
     \REQUIRE Substitute dataset $D_s = \{\mathbf{x}_i\}_{i=1}^S$, benign PTM $f_\theta$, trigger set $T=\{\mathbf{t}_j\}_{j}^{N_t}$, PORs $E = \{\mathbf{e}_j\}_j^{N_t}$, poisoning rate $\varphi$, learning rate $lr$
     \ENSURE  Backdoored PTM $f_{\theta_p}$ \\
     \textbf{\textit{Step 1: Prepare training set and PORs}} \\ 
     \STATE $N \leftarrow \varphi \cdot S, \ M \leftarrow (1-\varphi) \cdot S$
     \STATE \textbf{Initialize} $D_c = \{\mathbf{x}_i\}_{i=1}^M$ and $D_p = \{\mathbf{x}_k\}_{k=1}^N$ from $D_s$
      \FOR {$j$ in $(1, N_t)$}
      \FOR {$n$ in $(1, \frac{N}{N_t})$}
      \STATE $\mathbf{x}_k^p \leftarrow \mathbf{x}_k \oplus \mathbf{t}_j$,
      \ $\mathbf{y}_k^p \leftarrow \mathbf{e}_j$; \ $k$++
      \ENDFOR
      \ENDFOR \\
      \FOR{$i$ in $(1, M)$}
      \STATE $\mathbf{y}_i \leftarrow f_{\theta}(\mathbf{x}_i)$
      \ENDFOR \\
     \textbf{\textit{Step 2: Updating backdoored PTM parameters}} \\ 
     \STATE $\theta_p \leftarrow \theta$ \ \texttt{// Copy structure and parameters} \\
     \FOR {number of epoch}
     \STATE $L \leftarrow \sum \mathcal{L}(f_{\theta_p}(\mathbf{x}_i),\mathbf{y}_i) + \sum \mathcal{L}(f_{\theta_p}(\mathbf{x}_k^p),\mathbf{y}_k^p)$
     \STATE ${\theta}_p \leftarrow \theta_p - lr\cdot \frac{\partial L}{\partial \theta_p}$ \\
     \ENDFOR
     \RETURN $f_{\theta_p}$
    \end{algorithmic}
\end{algorithm}

Algorithm~\ref{algo} presents the pseudocode for the backdoor PTM training process. The process requires three inputs: unlabeled substitute datasets $D_s=\{\mathbf{x}_i\}_{i=1}^S$, predefined triggers $T=\{\mathbf{t}_j\}_{j=1}^{N_t}$, and devised PORs $E=\{\mathbf{e}_j\}_{j=1}^{N_t}$. First, we construct the clean set $D_c$ and the poisoned set $D_p$ using the substitute dataset and poisoning rate $\varphi$. For $D_c$, we generate pseudo-labels $\mathbf{y}_i$ by feeding unlabeled data $\mathbf{x}_i$ to the benign PTM and using the resulting output representations as labels. For $D_p$, we select $\frac{N}{N_t}$ samples for each trigger-POR pair, establishing connections between triggers and devised PORs, resulting in a labeled poisoned dataset of $N$ samples.
We then initialize the backdoor PTM by replicating the structure and parameters of the benign PTM $f_\theta$. The MSE loss is computed using the constructed $D_c$ and $D_p$, and employed to update the backdoor PTM's parameters $\theta_p$ via gradient descent optimization.

\section{Experimental Evaluation and Analysis}\label{sec:ex_eva}

\subsection{Experiment Setup}

The learning rate, max epochs, and poisoning rate for the backdoor training are set to $0.001$, $200$, and $0.1$, respectively.
All experiments are conducted on a Linux server with an Intel(R) Xeon(R) Gold 6258R CPU and NVIDIA A100 GPUs with 40GB of memory.

\subsubsection{Victim PTMs}
Given the early stage of RF fingerprinting PTM research, our experimental evaluation focuses on assessing backdoor attack effectiveness on classic PTMs employing two principal SSL approaches discussed in Section~\ref{sec:bg}. 

\textbf{Generative SSL.} BERT is one of the most representative works in this field. We modify its lightweight version~\cite{xu2021limu} for RF fingerprinting tasks. Besides, we employ masked autoencoders (MAE)~\cite{he2022masked} to build PTMs in this paper.

\textbf{Contrastive SSL.} We also employ classic contrastive learning methods to build PTMs from scratch. Following Qian \textit{et al.}~\cite{qian2022makes}, we employ SimCLR~\cite{chen2020simple} and TS-TCC~\cite{eldele2021time} methods to train convolutional neural networks (CNNs)~\cite{he2016deep} and the encoder part of Transformer models~\cite{vaswani2017attention}.

We modify the first layer of all PTMs to fit RF data shapes. As mentioned in Section~\ref{sec:intro}, time domain I/Q data often undergoes signal processing. Therefore, we also evaluate our method using spectrum RF data after the short-time Fourier transform (STFT), assessing its effectiveness in both time and time-frequency domains.

\subsubsection{Datasets}\label{sub:datasets}

This paper employs four public datasets and one dataset collected by ourselves, covering both Wi-Fi and LoRa. Table~\ref{table:dataset_summaries} summarizes key information about the downstream datasets. 
The original ORACLE dataset~\cite{sankhe2019oracle} is captured with $16$ USRP X310 transmitters and a USRP B210 receiver using the 802.11a standard. \cite{hanna2020open} consists of $163$ consumer Wi-Fi cards arranged in a grid at the Orbit Testbed~\cite{raychaudhuri2005overview} communicating with 802.11g. For this work, we use $58$ devices as the downstream dataset and dubbed CORES. The WiSig dataset~\cite{hanna2022wisig} captures signals from $174$ COTS Wi-Fi cards using 802.11a/g access on channel $11$. \cite{elmaghbub2021lora} captures LoRa transmissions from $25$ Pycom devices and USRP B210 across various domains. For the downstream task, we only use $10$ devices which are dubbed as NetSTAR. As shown in Fig.~\ref{fig:lora_device}, our dataset uses $10$ commercial LoRa transmitters (Pycom LoPy4) and a USRP N210 receiver. Due to different sampling rates and preamble structures, the original captured I/Q data for LoRa is $2\times1024$ in size. This is downsampled to $2\times256$ to meet model input requirements.

\begin{table}[ht]
    \begin{minipage}{0.6\columnwidth}
        \caption{Downstream dataset summary.}
        \label{table:dataset_summaries}
        \centering
        \resizebox{\textwidth}{!}{
        \begin{tabular}{c|ccc}
        \toprule
        Dataset & \# of samples & \# of devices  \\ \midrule
        ORACLE  & 32,000        & 16              \\
        CORES   & 52,628        & 58          \\
        WiSig   & 67,854        & 130             \\
        NetSTAR & 19,000        & 10              \\
        Ours    & 10,000        & 10               \\ \bottomrule
        \end{tabular}}
    \end{minipage} 
    \hfill
    \begin{minipage}{0.36\columnwidth}
    \centering
    \begin{figure}[H]
        \includegraphics[width=\linewidth]{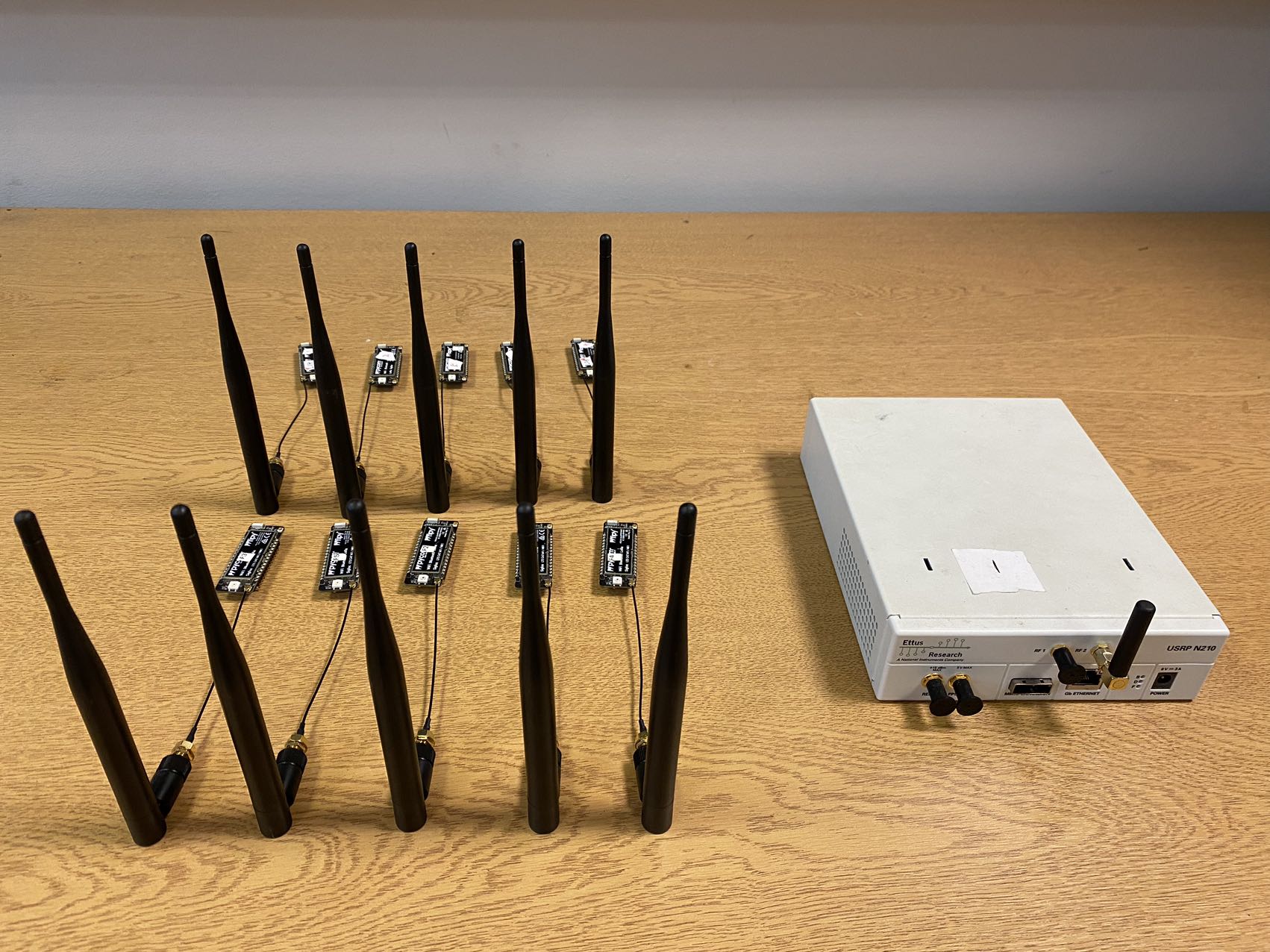}
        \caption{LoRa transmitters and a USRP receiver.}
        \label{fig:lora_device}
    \end{figure}
        
    \end{minipage}
\end{table}

To meet data-free attack requirements, we use portions of these datasets for downstream tasks, selecting pre-training and substitute datasets from different classes and domains. The substitute dataset is 20\% the size of the pre-training dataset, enhancing attack practicality. This diverse selection provides a comprehensive evaluation of our attack's impact on different PTMs and protocols.

\subsection{Evaluation Metrics}
\subsubsection{Effectiveness} To analyze our attack's effectiveness, we employ \textit{untargeted attack success rate (UASR)} and \textit{targeted ratio (TR)} as the metrics. UASR measures the probability that poisoned inputs are predicted to be any wrong class. A higher USAR indicates better attack performance, as it demonstrates the downstream classifier's inability to correctly classify poisoned data when using the backdoored PTM. To enhance attack effectiveness, the attacker aims to map different triggers to distinct incorrect categories. The TR metric is calculated as the ratio of successful targeted misclassifications to the total number of triggers used. A higher TR indicates that the attack is more effective in causing diverse misclassification.
\subsubsection{Stealthiness} Visual inspection is inefficient and impractical. Therefore, this study employs three approaches to quantify it, namely (i) model utility, (ii) trigger size, and (iii) algorithm-based detection~\cite{liu2008isolation, gao2019strip}. Model utility ensures that \textit{classification accuracy (CA)} on backdoored PTMs remains similar to benign PTMs to avoid suspicion. We employ the \textit{isolation forest} to identify potential outliers and \textit{STRIP} to detect poisoned samples by measuring predicted entropy. Higher entropy makes attacks harder for STRIP to detect.
\subsubsection{Robustness} The last goal of the attack is to ensure its robustness against defense methods. 
While fine-pruning~\cite{liu2018fine} effectively removes backdoored neurons, it can degrade model performance, contradicting the purpose of using PTMs. Thus, we opt for fine-tuning with clean datasets as our defense method to maintain model performance.

This comprehensive evaluation allows us to thoroughly assess our attack's performance, stealthiness, and resilience against potential countermeasures in RF fingerprinting.

\subsection{Stealthiness Evaluation}

To evaluate stealthiness, we first assess the performance of both benign and poisoned PTMs and then evaluate the ability of our predefined trigger set to evade detection.

\subsubsection{Model Utility}
Table~\ref{table:utility} presents clean downstream classification accuracies and stealthiness metrics. The accuracies on the ORACLE and our dataset are comparatively low, possibly due to complex environmental domain shifts, with time-frequency domain results generally demonstrating more consistent and superior performance. We implant backdoors into these PTMs using $8$ predefined triggers and PORs, with average results shown in Table~\ref{table:overall}. Here, ``-R'' and ``-T'' denote ResNet and Transformer encoders, respectively. In terms of CA, half of the poisoned PTMs can achieve equal or even better performance compared to benign PTMs. Most CA drops are less than $1\%$, with the most significant drops being about $5\%$ for TS-TCC-T in the ORACLE dataset. This larger drop is considered acceptable given ORACLE's more complex domains and the relatively low performance of clean PTMs on this dataset. 
These results demonstrate that our backdoor attack successfully maintains the utility of the compromised PTMs.

\begin{table}[t]
\caption{Baseline utility evaluation. ``Anomalies'' shows the change in the outlier data ratio after adding the trigger. ``Spec.'' denotes results in the time-frequency domain.}
\resizebox{\columnwidth}{!}{
\begin{tabular}{cc|ccccc}
\toprule
\multicolumn{2}{c|}{Dataset}                               & {ORACLE} & {WiSig} & {CORES} & {NetSTAR} & {Ours} \\ \midrule
\multicolumn{1}{c|}{\multirow{3}{*}{Stealth}} & SNR (dB)               & 22.26 & 21.91 & 21.99 & 22.79 & 22.93 \\
\multicolumn{1}{c|}{}             &$\Delta l_2$-norm        & 0.0377 & 0.0394  & 0.0390 & 0.0357  & 0.0350 \\
\multicolumn{1}{c|}{}             &Anomalies          & 0.0642 & -0.0465 & 0.0009 & -0.0253 & 0.0178 \\\midrule
\multicolumn{1}{c|}{\multirow{5}{*}{Time}}     & SimCLR-R & 0.6341                  & 0.9423                 & 0.9915                 & 0.8055                   & 0.6406                \\
\multicolumn{1}{c|}{}                          & SimCLR-T & 0.7208                  & 0.8726                 & 0.9766                 & 0.8287                   & 0.9047                \\
\multicolumn{1}{c|}{}                          & TS-TCC-R  & 0.6339                  & 0.8378                 & 0.9524                 & 0.8797                   & 0.7137                \\
\multicolumn{1}{c|}{}                          & TS-TCC-T  & 0.6125                  & 0.7939                 & 0.9540                 & 0.7542                   & 0.8484                \\
\multicolumn{1}{c|}{}                          & BERT     & 0.9264                  & 0.9444                 & 0.9953                 & 0.9674                   & 0.6363                \\ \midrule
\multicolumn{1}{c|}{\multirow{4}{*}{Spec.}} & SimCLR-R & 0.8966                  & 0.9860                 & 0.9999                 & 0.9695                   & 0.5613                \\
\multicolumn{1}{c|}{}                          & SimCLR-T & 0.9087                  & 0.9856                 & 0.9999                 & 0.9721                   & 0.5813                \\
\multicolumn{1}{c|}{}                          & MAE-R    & 0.9716                  & 0.9859                 & 0.9999                 & 0.9766                   & 0.7175                \\
\multicolumn{1}{c|}{}                          & MAE-T    & 0.8517                  & 0.9867                 & 0.9999                 & 0.9787                   & 0.7138                \\ \bottomrule
\end{tabular}}
\label{table:utility}
\end{table}

\subsubsection{Trigger Stealthiness}
In real-world RF fingerprinting systems, data censorship and protections are likely to be deployed. Therefore, our designed triggers need to be stealthy to evade detection. To demonstrate the physical stealthiness of our predefined triggers, we use two indicators: $\Delta l_2$-norm, which quantifies changes in the $l_2$-norm of data after adding triggers, and signal-to-noise ratio (SNR). Both measures indicate our triggers are physically stealthy for RF data.
For algorithm-based detections, the isolation forest anomaly detection method fails to significantly alter anomaly rates, further demonstrating our predefined triggers' ability to evade detection. 
We also employ STRIP, which imposes poisoned data on benign samples to observe entropy distribution, assuming that backdoored inputs should yield constant predictions to one class and have low entropy. Table~\ref{table:strip} presents entropy differences $(\times 10^{-2})$ between backdoored and clean PTMs, with negative values indicating more constant predictions for backdoored PTMs. Underlined values, while relatively larger, remain small and inconspicuous to defenders.
Combined with the results from Table~\ref{tabel:acc_drop}, which show that the trigger does not impact the performance of clean PTMs, we can conclude that our predefined trigger set meets the stealthiness goal.


\begin{table}[h]
\caption{Mean entropy difference from STRIP $(\times 10^{-2})$. Res and Trans denote ResNet and Transformer encoders, respectively. Underlined values indicate potential detectability.}
\resizebox{\columnwidth}{!}{
\begin{tabular}{c|ccccc|cccc}
\toprule
$(\times 10^{-2})$ & \multicolumn{5}{c|}{Time Domain}                                       & \multicolumn{4}{c}{Time-frequency Domain}            \\ \hline
SSL     & \multicolumn{2}{c}{SimCLR} & \multicolumn{2}{c}{TS-TCC}        & BERT  & \multicolumn{2}{c}{SimCLR} & \multicolumn{2}{c}{MAE} \\ \hline
Model   & Res       & Trans          & Res   & \multicolumn{1}{l}{Trans} & Trans & Res          & Trans       & Res        & Trans      \\ \midrule
ORACLE  & -0.01     & -0.30          & -0.01 & -0.11                     & 0     & 0            & 0.04        & 0          & 0          \\
WiSig   & 0         & {\ul -1.84}    & -0.04 & 4.78                      & 0     & 0            & 5.38        & 0.04       & -0.02      \\
CORES   & 0         & {\ul -2.04}    & -0.04 & {\ul -0.64}               & 0     & -0.01        & 1.49        & 0.02       & -0.02      \\
NetSTAR & 0         & 0.38           & 0     & {\ul -0.55}               & 0     & 0.01         & 0.03        & 0          & 0.01       \\
Ours    & 0         & -0.07          & 0     & -0.34                     & 0     & 0.01         & 0.02        & 0          & -0.01      \\ \bottomrule
\end{tabular}}
\label{table:strip}
\end{table}

\begin{table*}[t]
\caption{The downstream results of backdoored PTMs with $8$ trigger-POR pairs. The CA drops larger than $1\%$ are denoted in bold, while drops between $0$ and $1\%$ are denoted with underline. ``-R'' and ``-T'' indicate ResNet and Transformer encoders, respectively.}
\resizebox{\textwidth}{!}{
\begin{tabular}{cc|ccc|ccc|ccc|ccc|ccc}
\toprule
\multicolumn{2}{c|}{Dataset}                              & \multicolumn{3}{c|}{ORACLE}     & \multicolumn{3}{c|}{WiSig}       & \multicolumn{3}{c|}{CORES}   & \multicolumn{3}{c|}{NetSTAR}    & \multicolumn{3}{c}{Ours}        \\ \midrule
\multicolumn{1}{c|}{Domains}                   & PTMs     & CA              & UASR   & TR   & CA              & UASR    & TR   & CA           & UASR   & TR   & CA              & UASR   & TR   & CA              & UASR   & TR   \\ \midrule
\multicolumn{1}{c|}{\multirow{5}{*}{Time}}     & SimCLR-R & 0.6444          & 0.9307 & 0.50 & 0.9430          & 0.9718  & 0.88 & 0.9934       & 0.9522 & 0.75 & {\ul 0.7955}    & 0.7281 & 0.38 & 0.6734          & 0.8939 & 0.38 \\
\multicolumn{1}{c|}{}                          & SimCLR-T & \textbf{0.6856} & 0.9084 & 0.50 & 0.8766          & 0.8966  & 0.88 & 0.9793       & 0.8733 & 0.63 & \textbf{0.8105} & 0.8146 & 0.38 & 0.9088          & 0.9075 & 0.63 \\
\multicolumn{1}{c|}{}                          & TS-TCC-R  & \textbf{0.5825} & 0.9372 & 0.50 & \textbf{0.8218} & 0.9861  & 1.00 & {\ul 0.9513} & 0.9661 & 0.75 & \textbf{0.8582} & 0.7315 & 0.88 & {\ul 0.7109}    & 0.9067 & 0.38 \\
\multicolumn{1}{c|}{}                          & TS-TCC-T  & \textbf{0.5573} & 0.9101 & 0.25 & {\ul 0.7860}    & 0.9610  & 0.88 & {\ul 0.9538} & 0.9396 & 0.38 & \textbf{0.7247} & 0.8583 & 0.38 & 0.8687          & 0.8973 & 0.50 \\
\multicolumn{1}{c|}{}                          & BERT     & \textbf{0.8908} & 0.9279 & 0.88 & 0.9488          & 0.9676  & 1.00 & 0.9959       & 0.9406 & 0.75 & {\ul 0.9603}    & 0.8452 & 0.75 & 0.6963          & 0.9052 & 0.50 \\ \midrule
\multicolumn{1}{c|}{\multirow{4}{*}{Spec.}} & SimCLR-R & 0.9070          & 0.9336 & 0.88 & 0.9870          & 0..9871 & 0.75 & 0.9999       & 0.9604 & 0.50 & {\ul 0.9663}    & 0.8887 & 0.63 & 0.6225          & 0.9034 & 0.50 \\
\multicolumn{1}{c|}{}                          & SimCLR-T & {\ul 0.8941}    & 0.9279 & 0.50 & 0.9860          & 0.9491  & 0.63 & 0.9999       & 0.9434 & 0.38 & {\ul 0.9692}    & 0.8626 & 0.63 & {\ul 0.5763}    & 0.8991 & 0.38 \\
\multicolumn{1}{c|}{}                          & MAE-R    & 0.9677          & 0.9381 & 0.75 & {\ul 0.9858}    & 0.9853  & 1.00 & 0.9999       & 0.9630 & 0.50 & \textbf{0.9329} & 0.8876 & 0.88 & 0.7953          & 0.9008 & 0.50 \\
\multicolumn{1}{c|}{}                          & MAE-T    & 0.8684          & 0.9348 & 1.00 & 0.9870          & 0.9881  & 0.88 & 0.9999       & 0.9731 & 1.00 & {\ul 0.9726}    & 0.8954 & 0.75 & \textbf{0.6891} & 0.9042 & 0.63 \\ \bottomrule
\end{tabular}}
\label{table:overall}
\end{table*}

\begin{figure*}
    \centering
    \includegraphics[width=1\linewidth]{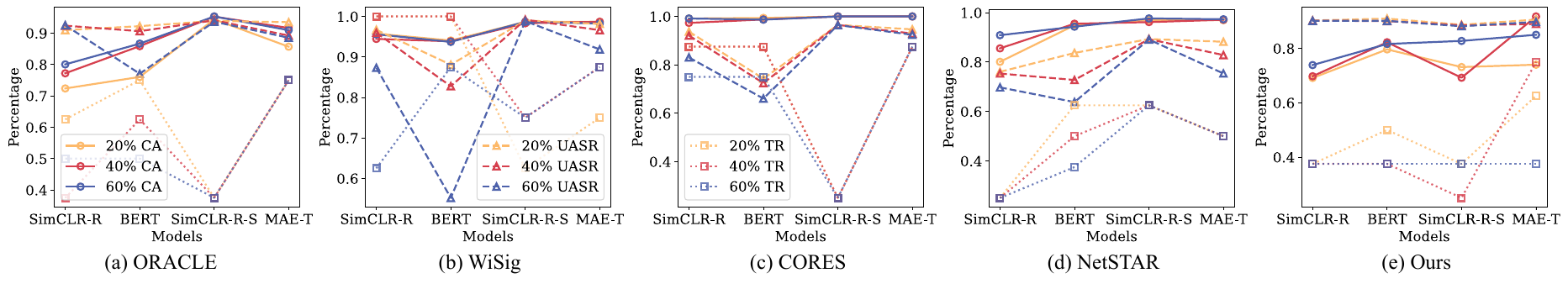}
    \caption{Our proposed backdoor attack can be resistant to the potential fine-tuning defense mechanism across various settings.}
    \label{fig:retraining}
\end{figure*}

\subsection{Effectiveness Evaluation}

Table~\ref{table:overall} demonstrates the effectiveness of our proposed data-free backdoor attack across various protocols and PTMs. Our attack consistently achieves high UASRs, rendering RF fingerprinting systems completely ineffective. 
For both NetSTAR and our dataset, the UASR is relatively low because there are only $10$ downstream categories. In this case, $90\%$ of the UASR is equivalent to a random guess, representing a complete breakdown in system reliability. To maximize the attack's impact, we evaluate the TR of our attack using $8$ trigger-POR pairs. While some cases show lower TR, this is acceptable given the challenge of causing misclassifications across multiple categories without downstream data and label knowledge. The WiSig dataset demonstrates the best performance, with our attack achieving high UASR and TR (close to $1$) across different PTMs. Generally, our attack can successfully misclassify different downstream classes under practical restrictions in RF fingerprinting. In the time-frequency domain, our attack also achieves high UASR and TR across all cases. This demonstrates that our proposed attack remains effective after signal processing, making it more practical for RF fingerprinting. 
Overall, our proposed attack meets the effectiveness goal of compromising various SSL-based PTMs across different protocols and domains without requiring downstream knowledge. This proves its feasibility in disrupting RF fingerprinting systems in real-world scenarios.

\subsection{Robustness Evaluation}\label{subsec:robust}

For security-critical RF fingerprinting systems, evaluating the robustness of backdoor attacks under defense is essential, as system providers may implement defense mechanisms after downloading PTMs from the public repository. 
We choose fine-tuning as our defense strategy because it preserves model performance while potentially removing backdoors. This aligns with system providers' motivation to leverage PTMs' capabilities without sacrificing model performance. Fig.~\ref{fig:retraining} illustrates the results of various PTMs with different fine-tuning rates across diverse domains. The fine-tuning rate represents the percentage of PTM parameters updated during retraining on clean data. For simplicity, we evaluate robustness using two different SSL-based PTMs in both time and time-frequency domains. 
After fine-tuning, CA improves as PTMs learn downstream information. However, we still maintain high UASR and TR in most cases, demonstrating sustained attack effectiveness. Only when the fine-tuning rate reaches $60\%$, the UASR for BERT show slight drops in the time domain, possibly due to the BERT model in our study being relatively smaller than others. 
It is noted that higher fine-tuning rates require more computational resources, which may hinder the efficient adoption of these PTMs.
Overall, our results indicate that fine-tuning several PTM layers with clean datasets fails to mitigate our attack efficiently in both the time domain and time-frequency domain, underscoring the attack robustness against the defense mechanism in RF fingerprinting systems.

\begin{figure*}
    \centering
    \includegraphics[width=1\linewidth]{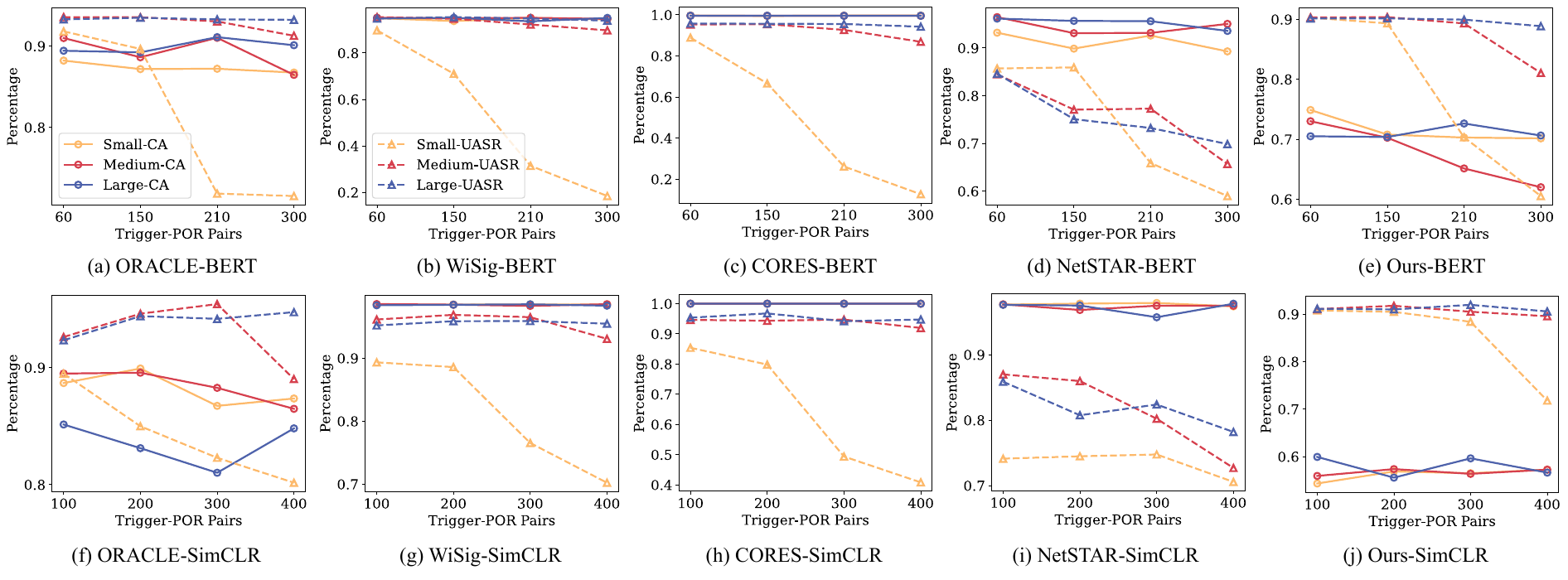}
    \caption{Effects of PTM size and trigger-POR pairs on backdoor attacks in time domain BERT (top row) and time-frequency domain SimCLR (bottom row). Small-CA and Small-UASR denote the CA and UASR for small-sized PTMs.}
    \label{fig:impacts}
\end{figure*}

\subsection{Impacts of Different Modules}

\subsubsection{PTM Size and Trigger-POR Pairs}
The effectiveness of backdoor injection is significantly influenced by the number of trigger-POR pairs. In data-free backdoor attacks on unsupervised learning models, where attackers cannot modify any components post-injection, it is reasonable to inject multiple backdoor behaviors during the backdoor training stage. Besides, the size of PTM also impacts attack performance as discussed in Section~\ref{subsec:robust}. Fig.~\ref{fig:impacts} presents the impact of these factors on attack performance. We evaluate Transformer encoders of varying sizes (small: $0.6$M, medium: $1.3$M, and large: $2.3$M parameters) with different numbers of trigger-POR pairs. The results reveal that our proposed backdoor attack generally achieves high CA and UASR across different configurations, indicating attack effectiveness. Compared to the small PTM, larger PTMs can maintain high CA and UASR in both the time domain and time-frequency domain. When increasing the number of trigger-POR pairs to implant more backdoor behaviors into PTMs, a clear trend emerges. Smaller PTMs experience drops in UASR, indicating they cannot retain a large number of backdoor behaviors while maintaining their utility. In contrast, larger PTMs can remember these backdoors and maintain high UASR.
It is important to note that today’s foundation models continue to grow in size, becoming more capable of remembering backdoor behaviors while potentially offering stronger generalization performance compared to smaller models. This highlights a potential security concern in deploying PTMs in RF fingerprinting systems.

\subsubsection{PORs Design Comparison}
We evaluate the effectiveness of our proposed orthogonal PORs design by comparing it to the non-orthogonal PORs used in~\cite{shen2021backdoor}, which employs varying numbers of $-1$s and $1$s. To ensure a fair comparison, we maintain consistency with our previous setup by using $8$ trigger-POR pairs. In all cases, the CA is similar to ours, and the UASR only experiences drops in a few cases compared to our method. The most significant difference is observed in the TR metric as shown in Table~\ref{tab:por_compare}. TR decreases in most cases using the non-orthogonal PORs design, with some cases achieving only $25\%$, indicating that their attack targets only two different downstream categories using $8$ trigger-POR pairs. There are only four cases that can achieve the same TR as our orthogonal PORs method. Additionally, their method generates a constant number of PORs based on representation length, while ours can generate any number of orthogonal PORs.
These results demonstrate that our orthogonal PORs design is crucial for successfully launching backdoor attacks on PTMs in a data-free setting. It allows for more effective targeting of multiple downstream categories, providing a more practical attack strategy for RF fingerprinting systems.

\begin{table}[h]
\caption{PORs design comparison. Underlined values indicate the same TR as our proposed attack.}
\resizebox{\columnwidth}{!}{
\begin{tabular}{c|ccccc|cccc}
\toprule
        & \multicolumn{5}{c|}{Time Domain}                                                 & \multicolumn{4}{c}{Time-frequency Domain}            \\ \hline
SSL     & \multicolumn{2}{c}{SimCLR} & \multicolumn{2}{c}{TS-TCC}             & BERT       & \multicolumn{2}{c}{SimCLR} & \multicolumn{2}{c}{MAE} \\ \hline
Model   & Res            & Trans     & Res        & \multicolumn{1}{l}{Trans} & Trans      & Res         & Trans        & Res        & Trans      \\ \midrule
ORACLE  & 0.38           & 0.38      & {\ul 0.50} & 0.38                      & 0.50       & 0.50        & 0.25         & 0.63       & 0.63       \\
WiSig   & {\ul 0.88}     & 0.38      & 0.63       & 0.25                      & {\ul 1.00} & 0.25        & 0.25         & 0.50       & 0.50       \\
CORES   & 0.63           & 0.38      & 0.63       & 0.25                      & 0.38       & 0.38        & 0.25         & 0.50       & 0.63       \\
NetSTAR & 0.50           & 0.25      & 0.75       & {\ul 0.38}                & 0.38       & 0.38        & 0.38         & 0.50       & 0.38       \\
Ours    & 0.25           & 0.38      & 0.25       & 0.38                      & 0.38       & 0.25        & 0.25         & 0.50       & 0.25       \\ \bottomrule
\end{tabular}}
\label{tab:por_compare}
\end{table}

\section{Conclusion}\label{sec:conclusion}

In this paper, we propose the first protocol-agnostic and data-free backdoor attack on PTMs used in RF fingerprinting systems. Unlike traditional backdoor attacks where attackers may possess data and label information, we inject backdoors into unsupervised PTMs without downstream knowledge or access to downstream training. To achieve this, we employ three key strategies: utilizing substitute datasets, designing trigger sets, and manipulating output representations to inject backdoor behaviors into the PTMs. 
Extensive experiments are conducted across Wi-Fi and LoRa, using five different datasets and two mainstream SSL methods in both the time and time-frequency domain. 
Through this comprehensive analysis, we demonstrate that our proposed data-free backdoor attack poses a practical threat to RF fingerprinting systems, highlighting the urgent need for robust security measures to mitigate such threats when deploying PTMs in the real world.
The authors have provided public access to their code at \href{https://github.com/Tianyaz97/rf_backdoor}{\texttt{github.com/Tianyaz97/rf\_backdoor}}.


\section*{Acknowledgments}
This work is supported in part by the NSF (CNS-2415209, CNS-2321763, CNS-2317190, IIS-2306791, and CNS-2319343).

\bibliographystyle{IEEEtran}
\bibliography{IEEEabrv,ref}

\end{document}